# An Exchange Rate Target Zone Model with a Terminal Condition and Mean-Reverting Fundamentals

**Viktors Ajevskis**[1]
**Bank of Latvia**[2]

**Abstract.** This paper proposes a target zones exchange rate model with a terminal condition of entering a currency zone. It is assumed that the exchange rate is a function of the fundamental and time. Another essential assumptions of the model is that the fundamental process is bounded inside the band [-f,f] and that terminal condition for the exchange rate holds. Using Ito's lemma, we obtain a parabolic partial differential equation for the exchange rate. The fundamental is specified in two ways: as a regulated Brownian motion and Ornstein-Uhlenbeck processes. For the case of the Brownian motion process the closed form solution of the problem is obtained, whereas for the Ornstein-Uhlenbeck process the closed form solution does not exist, therefore we had to use numerical method for solving of the problem. Both specifications are compared numerically.

Key words: Ornstein-Uhlenbeck process, Ito's lemma, Kumer function

## 1.INTRODUCTION

One of the most important problems in macroeconomics is developing of currency exchange rate models. Models help to understand exchange rate dynamics as well as interaction of an exchange rate with other macroeconomic variables. There are two main types of exchange rate regimes: free-floating and fixed to some currency or currency basket. So-called target zones are hybrid of the two. When maintaining target zone policy, monetary authorities commit themselves to keep an exchange rate to definite currency or basket of currencies inside definite bands around some parity value.

The most successful exchange rate target zone model was proposed by P.Krugman (see Krugman, P. (1991)). Krugman has assumed, that the exchange rate, similarly to the prices of other assets, depends linearly on aggregated "fundamental", including various fundamental determinants of a currency rate (such as the domestic output, the money supply, the foreign interest rates, the price levels etc.) and on the expected change in future exchange rate. The fundamental contains two components: exogenous and stochastic component, called "velocity" and money supply, controlled by the central bank. Krugman assumed that velocity is Brownian motion without a drift. In his model Krugman substantially used assumption that target zone exists forever. This assumption is not realistic, because a target zone typically exists finite period of time.

[1] email:Viktors.Ajevskis@bank.lv
[2] The views expressed in this paper are the sole responsibility of the author and do not necessarily reflect the position of the Bank of Latvia.

In our work we generalize Krugman's model for the case when target zone exists finite time, taking into account a terminal condition on exchange rate. An example of such situation is ERM II, which suppose that countries participating in this mechanism have to keep ±15% wide target zone against euro. After successful fulfillment of the latter and other Maastriht criterions, these countries will abandon national currencies and will enter euro currency zone. In our previous research (Ajevskis 2011) we specified fundamental as regulated Brownian motion. In the preset article we consider fundamental specified as regulated Ornstein-Uhlenbeck process. Such a specification corresponds to the monetary policy that allows marginal as well as intramarginal interventions. Unfortunatelly, for the Ornstein-Uhlenbeck fundamental an analytical solution of non-stationary problem does not exist, therefore we had to use numerical method for solving of the problem. The results were compared with obtained for Brownian motion specified fundamental case.

2. A Target Zone Exchange Rate Model with a Terminal Condition of Entering Currency Zone and the Fundamental driven by Ornstein-Uhlenbeck Stochastic Process

The standard monetary exchange rate model was employed ([4]). In the model it is assumed that exchange rate $e(\tau)$ depends on fundamental $f(\tau)$, and on expected change in the exchange rate:

$$e(\tau) = f(\tau) + \alpha E_\tau[de(\tau)/d\tau], \quad \alpha > 0, \qquad (1)$$

where $E_\tau[de(\tau)/d\tau] = \lim_{s \to 0+} \frac{E_\tau[e(\tau+s)] - e(\tau)}{s}$,

and $E_\tau$ is the mathematical expectation operator conditioned on the information available up to time $\tau$.

Assume that foreign exchange market interventions, which directly affect the money supply, are undertaken to prevent the fundamental to move outside a specified band for the fundamental. Hence we assume that there exist lower and upper bounds for the fundamental, such that the fundamental fulfills

$$\underline{f} < f(\tau) < \overline{f}.$$

Marginal interventions are modelled using "Regulators" U and L, which are are monotonous continuous processes; marginal interventions can be represented by:

$$dm = dL - dU,$$

where $dL \geq 0$, $dU \geq 0$. $dL$ represents increase in money supply in case $f = \underline{f}$, and $dU$ represents reductions in money supply in case $f = \overline{f}$. Once the fundamental moves inside the band, marginal interventions cease. Assume now, that monetary policy allows also intramarginal interventions. This case can be modeled, using mean-reverting (Ornstein-Uhlenbeck) process

When fundamentals follow regulated Ornstein-Uhlenbeck process, stochastic differential equation for fundamental has the following form:

$$df(\tau) = -\rho(f(\tau) - \mu)d\tau + \sigma dw(\tau) - dU + dL, \qquad (2)$$

where $\rho$ and $\sigma$ are positive constants; $\mu$ is the long-run level of $f$ and $\rho$ is the speed of adjustment of the process towards $\mu$; $w(\tau)$ is a standard Wiener process with $E_\tau(dw) = 0$ and $E_\tau(dw^2) = d\tau$.

Now consider that function $e : R_+ \times R \to R$ is continuously differentiable with respect to the first argument and twice continuously differentiable with respect to the second argument. Applying Itô's lemma, we obtain:

$$de = \left( \frac{\partial e}{\partial \tau} - \rho(f(\tau) - \mu)\frac{\partial e}{\partial f} + \frac{\sigma^2}{2}\frac{\partial^2 e}{\partial f^2} \right)d\tau + \frac{\partial e}{\partial f}\sigma\, dw \qquad (3)$$

Taking mathematical expectation from both sides of equation (3) and noting that $E\tau(dw)=0$, we have:

$$E_\tau(de) = \left(\frac{\partial e}{\partial \tau} - \rho(f(\tau)-\mu)\frac{\partial e}{\partial f} + \frac{\sigma^2}{2}\frac{\partial^2 e}{\partial f^2}\right)d\tau.$$

Which imply:

$$\frac{E_\tau(de)}{d\tau} = \frac{\partial e}{\partial \tau} - \rho(f(\tau)-\mu)\frac{\partial e}{\partial f} + \frac{\sigma^2}{2}\frac{\partial^2 e}{\partial f^2}. \tag{4}$$

Substituting (4) into equation (1) gives

$$e = f(\tau) + \alpha\left(\frac{\partial e}{\partial \tau} - \rho(f(\tau)-\mu)\frac{\partial e}{\partial f} + \frac{\sigma^2}{2}\frac{\partial^2 e}{\partial f^2}\right).$$

By re-arranging the equation terms we get

$$\frac{\partial e}{\partial \tau} - \rho(f(\tau)-\mu)\frac{\partial e}{\partial f} + \frac{\sigma^2}{2}\frac{\partial^2 e}{\partial f^2} - \frac{1}{\alpha}e = -\frac{1}{\alpha}f(\tau). \tag{5}$$

Suppose that at some definite moment T in the future a country, which maintained exchange rate target zone regime, will enter currency zone of pegging currency. In such a case at time T exchange rate doesn't depend on f. This means, that the terminal condition:

$$e(T,f) = 0 \tag{6}$$

holds. It is obvious, that this condition is necessary to avoid arbitrage opportunity at the moment of entering the currency zone.

We apply transformation $t = T - \tau$ and rewrite the equation (5) in the form:

$$\frac{\partial e}{\partial t} + \rho(f(t)-\mu)\frac{\partial e}{\partial f} - \frac{\sigma^2}{2}\frac{\partial^2 e}{\partial f^2} + \frac{1}{\alpha}e = \frac{1}{\alpha}f(t) \tag{7}$$

Condition (6) will change to
$$e(0,f) = 0. \tag{8}$$

smooth pasting conditions are not affected by the transformation:

$$\frac{\partial e}{\partial f}(t,\underline{f}) = \frac{\partial e}{\partial f}(t,\bar{f}) = 0 \qquad (9)$$

The system (7)-(9) is an initial-boundary problem for the parabolic equation. We will seek for solution in the form

$$e(t,f) = e^*(t,f) + \hat{e}(f), \qquad (10)$$

where $\hat{e}(f)$ is a solution to the following stationary problem:

$$\frac{\alpha\sigma^2}{2}\frac{\partial^2 \hat{e}(f)}{\partial f^2} - \alpha\rho(f-\mu)\frac{\partial \hat{e}(f)}{\partial f} - \hat{e}(f) = -f, \qquad (11)$$

$$\frac{\partial \hat{e}}{\partial f}(\underline{f}) = \frac{\partial \hat{e}}{\partial f}(\bar{f}) = 0 \qquad (12)$$

It is not difficult to obtain the general solution of the stationary equation (11):

$$\hat{e}(f) = C_1 M\left(\left[\frac{1}{2\alpha\rho}\right],\left[\frac{1}{2}\right],\frac{\rho(\mu-f)^2}{\sigma^2}\right) +$$

$$+ C_2 \frac{\sqrt{\rho}(\mu-f)}{\sigma} M\left(\left[\frac{1+\alpha\rho}{2\alpha\rho}\right],\left[\frac{3}{2}\right],\frac{\rho(\mu-f)^2}{\sigma^2}\right) + \frac{\alpha\rho\mu + f}{1+\alpha\rho}$$

where $M([a],[b],[z])$ is the confluent hypergeometric Kummer function (Abramowitz, M., Stegun, I.A.(1964)); $C_1$ and $C_2$ are integration constants.

For symmetric case (when $\mu=0$ and $\underline{f} = -\bar{f}$) the constant $C_1=0$ and the values of $C_2$ and $\bar{f}$ can be obtained from the system of two equations:

$$\begin{cases} \dfrac{\bar{f}}{1+\alpha\rho} - C_2 \dfrac{\sqrt{\rho}\bar{f}}{\sigma} \cdot M\left(\left[\dfrac{1+\alpha\rho}{2\alpha\rho}\right],\left[\dfrac{3}{2}\right],\dfrac{\rho\cdot\bar{f}^2}{\sigma^2}\right) = \bar{e} \\ \dfrac{1}{1+\alpha\rho} - C_2 \dfrac{2}{3}\dfrac{\sqrt{\rho}(1+\alpha\rho)\bar{f}^2}{\alpha\sigma^3} \cdot M\left(\left[\dfrac{1+3\alpha\rho}{2\alpha\rho}\right],\left[\dfrac{5}{2}\right],\dfrac{\rho\cdot\bar{f}^2}{\sigma^2}\right) - \\ -C_2 \dfrac{\sqrt{\rho}}{\sigma}\cdot M\left(\left[\dfrac{1+\alpha\rho}{2\alpha\rho}\right],\left[\dfrac{3}{2}\right],\dfrac{\rho\cdot\bar{f}^2}{\sigma^2}\right) = 0 \end{cases} \qquad (13)$$

## 3. Numerical modelling

For the non-stationary problem (7)-(9) analytical solution does not exist. The solution can be obtained only by numerical methods, taking into account boundary values of fundamental, obtained from (13). For this purpose we used package Mathlab7.0. The numerical solution of the system (7)-(9) is shown on the figure 1. The system was solved for parameter values μ=0, ρ =1, σ = 0.1, α = 3, $\bar{e} = 0.01$ and $\underline{e} = -0.01$ (which corresponds to ±1% target zone about parity value).

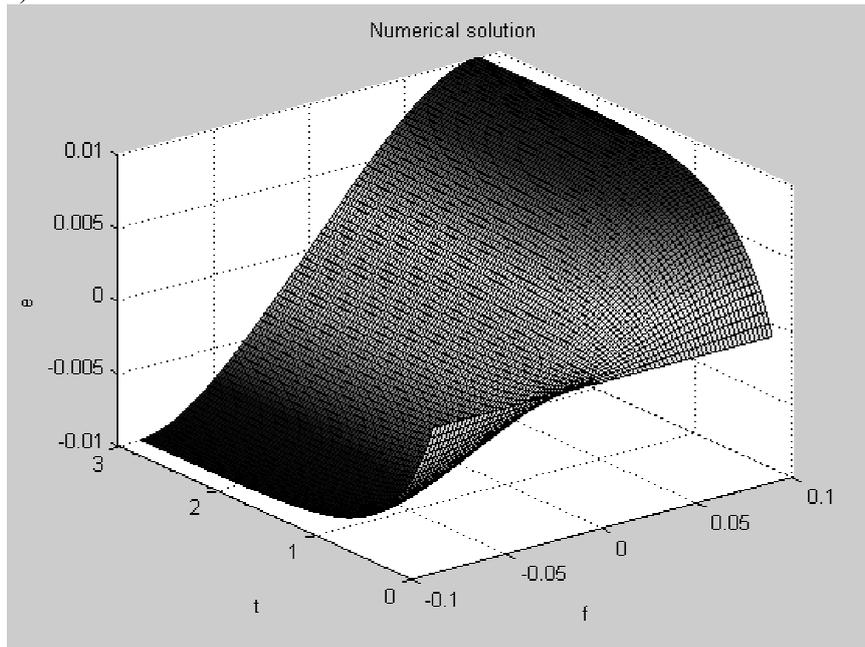

Figure 1. Exchange rate as a function of fundamental and time.

Figure 2 shows sections $e(\bullet, f)$ of the solution at times $t = 0, 0.15, 0.6, 1.95, 3$ years. As the Figure shows, the shorter period is left till entering currency zone, the less sensitive becomes exchange rate to fundamental changes. $t=3$ corresponds to solution of the Krugman's model.

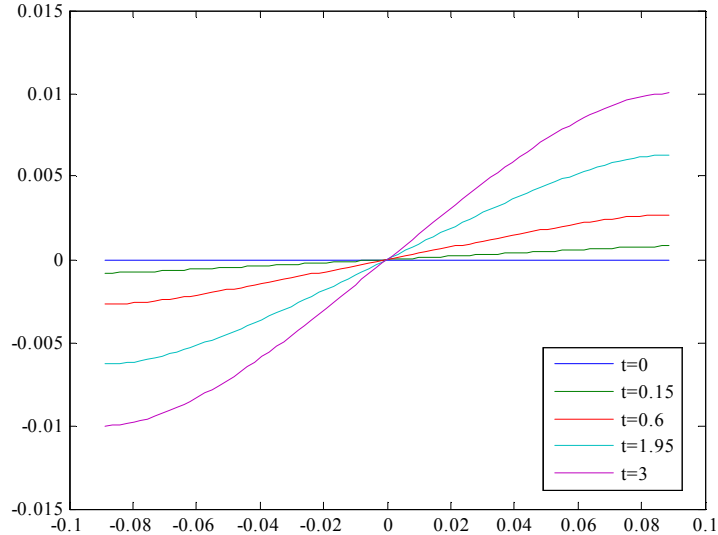

Figure 2. Sections of exchange rate as a function of fundamental for different time moments.

The solutions $e(t, \pm \bar{f})$, which correspond to boundary values, are plotted in Figure 3. The given solutions define the region, where exchange rate should be kept. The upper margin decreases and lower margin increases when the moment of entering the currency zone approaches. Absolute value of both margins decrease gradually and after 3 years (euro introducing time) converges to 0.

For comparison purpose we plotted together on one graph solutions for regulated Brownian motion and mean-reverting fundamentals. For small target zones there is small difference between exchange rates functional dependence on fundamental for the two specifications of the model. For the same values of parameters, model with mean-reverting fundamentals allow for slightly wider bands for fundamental than model with regulated Brownian motion fundamental.

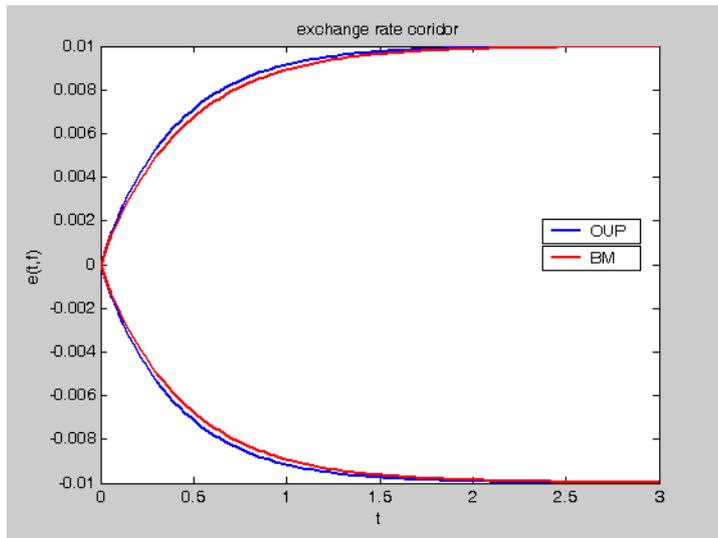

Figure 3. Time dynamics for the boundaries of the exchange rate target zone.

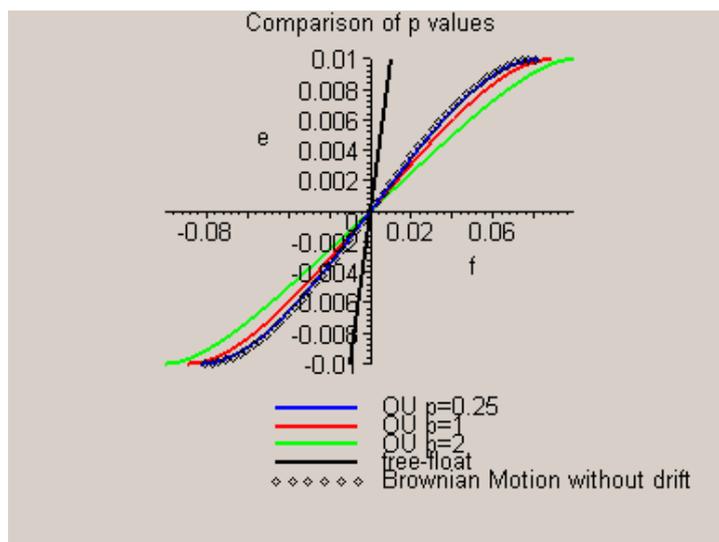

Figure 4. Solutions for stationary exchange rate target zone problem for the Brownian motion and OU processes with different persistence.

Figure 4 shows solutions of the stationary problem for different $\rho$ values. The straight line corresponds to free-float solution. The more intensive are intramarginal interventions, the wider is fundamental range for fixed width of target zone. As $\rho$ approaches 0, solution of the model with mean-reverting fundamental converges to the solution of the model with Brownian-motion specified fundamental.

CONCLUSIONS

The study presents a generalization of Krugman's Target Zones model for the case of mean-reverting fundamental and the presence of the terminal condition corresponding to joining a currency area.

The main assumptions of the model are that a central bank does not allow the fundamentals to exceed the interval $[-\bar{f}, \bar{f}]$ bounds, *"smooth pasting conditions"*, and zero terminal condition. The model with fundamental, followed by a regulated Ornstein-Uhlenbeck process corresponds to the monetary policy that allows a central bank caries out intramarginal interventions. For this case only stationary problem has analytical solution; for the non-stationary problem an analytical solution does not exist, therefore we have to use numerical method for solving the problem. The proposed model can be used by monetary authorities, who maintain a target zone currency policy and intend to enter a currency zone.

# REFERENCES

Abramowitz, M., Stegun, I.A.(1964): Handbook of mathematical function with Formulas, Graphs, and Mathematical Tables. Washington: U.S. Government Printing Office, 1964. P. 504-535.

Ajevskis, V. (2011): A target zone model with the terminal condition of joining a currency area". *Applied Economics Letters*. 2011, volume 18, number 13-15, 1273-1278.

Dumas, B.(1991): "Super Contact and Related Optimality Conditions", Journal of Economic Dynamics and Control, 15, 675–85.

Harrison, J. M.(1985): Brownian Motion and Stochastic Flow Systems. John Wiley & Sons, Inc. 1985.

Krugman, P.(1991): Target Zones and Exchange Rate Dynamics. Quarterly Journal of Economics. 106, pp 311-325. 1991.

Skohokhod, A.V., Gihman, I.I.(2004): The theory of stochastic processes. Springer Verlag, Berlin, Vol 1,2. 2004.